\documentclass[conference,letterpaper]{IEEEtran}

\usepackage[utf8]{inputenc} 
\usepackage[T1]{fontenc}
\usepackage{url}
\usepackage{ifthen}
\usepackage{cite}
\usepackage[cmex10]{amsmath} 
\usepackage{graphicx}
\usepackage{dcolumn}
\usepackage{bm}
\usepackage{hyperref}

\usepackage{booktabs}       
\usepackage{amsfonts}       
\usepackage{nicefrac}       
\usepackage{microtype}      
\usepackage{xcolor}         
\usepackage{amssymb}  
\usepackage{epsfig}
\usepackage{epstopdf}
\usepackage{mathtools}
\usepackage{etoolbox}
\usepackage[caption=false,font=footnotesize]{subfig}
\usepackage{soul}
\usepackage{verbatim}
\usepackage{algorithm,algorithmic}
\usepackage{bbm}

\newcounter{relctr} 
\everydisplay\expandafter{\the\everydisplay\setcounter{relctr}{0}} 

\AtBeginDocument{} 

\DeclareMathOperator\tr{tr}

\DeclareMathOperator{\tv}{TV}

\newcommand{\tminus}{\scalebox{0.9}[1.0]{\( - \)}}
\newcommand{\tplus}{\scalebox{0.9}[1.0]{\( + \)}}

\newtheorem{theorem}{\bf Theorem}

\newtheorem{definition}{\bf Definition}

\newtheorem{example}{\bf Example}

\begin{document}
\title{A Causal Model for Quantifying Multipartite Classical and Quantum Correlations} 


\author{%
  \IEEEauthorblockN{Shuchan Wang and Gerhard Wunder}
  \IEEEauthorblockA{Department of Mathematics and Computer Science\\ 
                    Freie Universität Berlin
                    \\Berlin, Germany\\
\{s.wang, g.wunder\}@fu-berlin.de}
}


\maketitle


\begin{abstract}
We give an operational definition of information-theoretic resources within a given multipartite classical or quantum correlation. We present our causal model that serves as the source coding side of this correlation and introduce a novel concept of resource rate. We argue that, beyond classical secrecy, additional resources exist that are useful for the security of distributed computing problems, which can be captured by the resource rate. Furthermore, we establish a relationship between resource rate and an extension of Shannon’s logarithmic information measure, namely, total correlation. 
\end{abstract}

\section{Introduction}
Quantum entanglement and its connections to classical information have been the subject of continuous research, finding applications in diverse fields such as quantum key distribution (QKD), superdense coding, and distributed computing \cite{horodecki2015axiomatic,acin2004multipartite,leskovjanova2020classical,scarani2009security,scarani2006secrecy,christandl2007unifying,augusiak2009multipartite,horodecki2009quantum,buhrman2010nonlocality,barrett2005nonlocal,masanes2006general,acin2005quantum}. 
One approach within QKD is entanglement-based QKD \cite{PhysRevLett.67.661}, where quantum entanglement is harnessed to achieve non-local effects and replace classical information. Especially in cryptographic tasks involving more than two parties, such as multipartite key distribution or conference key agreement (CKA), multipartite entanglement is utilized in different protocols \cite{hillery1999quantum, pivoluska2018layered,grasselli2018finite, proietti2021experimental} and leads to a higher secret key rate than bipartite protocol \cite{epping2017multi}. Recent advances in fully connected quantum networks \cite{wengerowsky2018entanglement,joshi2020trusted} further facilitate the realization of these protocols.

The quantification of secrecy in multipartite cryptography poses a persistent challenge. Classical information quantification of a given probability, pioneered by Shannon in his seminal work \cite{shannon_mathematical_1948}, is inherently linked to logarithmic measures such as entropy and relative entropy. 
To measure how much secrecy can be provided by a classical or quantum correlation, Cerf et al. introduce the concept of secrecy monotone in \cite{cerf2002multipartite}. Total correlation, an extension of Shannon's logarithmic measures introduced by Watanabe \cite{watanabe1960information}, serves as one of these secrecy monotones. Its operational significance extends to various aspects, including the variational form of the optimal broadcast rate of the omniscient coordination problem \cite{kurri2021coordination} and the multivariate covering lemmas  \cite{el2011network,yassaee2014achievability}. 
Recent developments in device-independent QKD provide quantum versions of total correlation, offering an upper bound on the rate of CKA  
\cite{winczewski2022limitations,horodecki2022fundamental,philip2023multipartite}.

Meanwhile, the discovery of cause-and-effect relationships with a given correlation has been an important task in causality \cite{pearl2009causality}. The problem of finding the minimal amount of resources to replicate a given effect, known as model preference and minimality in the context of causality, has also been extensively studied.  This includes the study of the coordination problem in classical information theory \cite{han1993approximation,cuff2010coordination,cuff2013distributed,yassaee2014achievability,kurri2021coordination} and the simulation of quantum measurements and quantum channels in quantum information theory \cite{winter2004extrinsic,bennett2014quantum,wilde2012information,atif2021faithful}. Additionally, an intriguing causal perspective of quantum mechanics called indefinite causal order has been studied in \cite{oreshkov2012quantum,goswami2018indefinite, oreshkov2016causal,abbott2016multipartite,rubino2017experimental}. The equivalence between positive secrecy content and quantum correlation has been established in \cite{acin2005quantum,masanes2006general,barrett2005nonlocal}.

In this paper, the classical and quantum correlations we study refer to joint probabilities and entangled states, respectively. Our main focus is on understanding the information-theoretic resources offered by a given multipartite classical or quantum correlation, and we provide its semantic and formal definitions later in the paper. Following Reichenbach’s Common Cause  Principle (RCCP) \cite{sep-physics-Rpcc} and no-signaling condition, we introduce a causal model where the correlation is replicated by a set of common causes. These common causes are considered as mediums through which messages can be transmitted securely, allowing us to discover the resources that cannot be measured by the classical secrecy measure in multipartite cryptographic tasks.

\subsection{Notations}  
$\mathbb{N}_+$ denotes the set of positive natural numbers.  $ \mathbb {F}_2$ is the finite field with two elements. Let $\mathcal{X}$ be a discrete finite set.   $\mathcal{X}^n$ denotes the $n$-ary Cartesian power set of $\mathcal{X}$. We write an element $x\in\mathcal{X}$ in lower-case letters and a random variable $X$ on $\mathcal{X}$ in capital letters. 
$X_{1:m}$ and $X^n_{1:m}$ denote sequences of random variables with the sizes of $m$ and $m\times n$. $X^{\otimes n}$ denotes a $n$-length sequence of i.i.d. random variables whose probability measure is given by the tensor product $\bigotimes^n_{i=1} P_{X_i}$. 
 $H(\cdot)$, $I(\cdot;\cdot)$ and $\tv(\cdot,\cdot)$ denote entropy, mutual information, and total variation, respectively. $\delta_{x}$ is a Dirac measure at $x$. All the logarithms are the logarithms to the base 2. 
 The occurrence of a sequence of events $A_n$ almost surely means that the probability $\lim_{n\rightarrow \infty}P(A_n) = 1$. $\epsilon$ and $\delta$ denote positive infinitesimal terms converging to zero when  $n \rightarrow \infty$. 
 Let $\mathcal{H}$ be a finite-dimensional Hilbert space. $\mathcal{S}(\mathcal{H})$ denotes the set of the density operators over $\mathcal{H}$. 
\section{Preliminaries}\label{sec:model}
\begin{figure}
    \centering
    \includegraphics{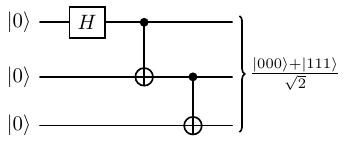}
    \caption{Quantum circuit of 3-qubit GHZ state.}
    \label{fig:circ1}
\end{figure}

This section gives an overview of multipartite quantum cryptography and secrecy monotone as well as a special case that supports our argument.
Current studies on multipartite quantum cryptography primarily focus on Greenberger-Horne-Zeilinger (GHZ) state \cite{greenberger1989going}, which is a fundamental multipartite entangled state. The $m$-qubit GHZ state in the computational basis is written as 
\begin{equation}\label{eq:Q1}
|\psi_1\rangle := \frac{1}{\sqrt{2}}(|0\rangle^{\otimes m}+|1\rangle^{\otimes m}).
\end{equation}
The corresponding quantum circuit is shown in Fig.~\ref{fig:circ1}. A $m$-qubit GHZ state exhibits instantaneous identical outcomes in $m$ subsystems when measured in the computational basis. The outcome is written as a uniform probability on two vectors $\boldsymbol{0}$ and $\boldsymbol{1}$ with all entries of $0$ and $1$ respectively, i.e.,  $P^{(1)}_{X_{1:m}} := \frac{1}{2}(\delta_{\boldsymbol{0}}+\delta_{\boldsymbol{1}})$. It is known as CKA and allows a user in the system to broadcast one perfect secure bit with the remaining $m-1$ users. The key with layered structure discussed in \cite{pivoluska2018layered} is a tensor product of different GHZ states among different subsets of the whole set of users.

\begin{figure}[ht]
         \centering
    \includegraphics[width=0.32\textwidth]{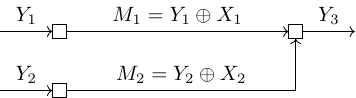}
\caption{A problem of distributed coding for computing.}
\label{fig:LatentStr3}
\end{figure}

Next, we recall quantum secret sharing in \cite{hillery1999quantum}. Define $|\psi_2\rangle$ as the 3-qubit GHZ state in the Hadamard basis, i.e.,
\begin{align}
|\psi_2\rangle :=& \frac{\sqrt{2}}{2}(|\tplus\tplus\tplus\rangle+|\tminus\tminus\tminus\rangle)\nonumber\\=& \frac{1}{2}(|000\rangle+|011\rangle+|101\rangle+|110\rangle).\label{eq:Q2}
\end{align}
The measurement outcome in the computational basis, i.e. $P^{(2)}_{X_{1:3}}:= \frac{1}{4}(\delta_{(000)}+\delta_{(011)}+\delta_{(101)}+\delta_{(110)})$,  yields $X_1, X_2 \sim \mathrm {Bernoulli} (\frac{1}{2})$ as two independent random variables, while deterministic relationship $X_3 = X_1 \oplus X_2 $ holds when all three are observed. 
Notably, these keys can be useful in a problem of distributed coding for computing \cite{cerf2002multipartite}, as shown in Fig. ~\ref{fig:LatentStr3}. The objective is to securely compute $Y_3 = Y_1\oplus Y_2$ at the receiver, where $Y_1$ and $Y_2$ are distributed binary sources available at two different senders. $Y_3$ can be recovered from two encrypted bits $Y_3 = M_1\oplus M_2 \oplus X_3$, while the original sources $Y_1$ and $Y_2$ remain private. We can generalize \eqref{eq:Q2} into higher dimension with $m\geq 3$. Let
\begin{equation*}
|\psi_3\rangle :=\frac{1}{\sqrt{2}}(|\tplus\rangle^{\otimes m}+|\tminus\rangle^{\otimes m})= 2^{-\frac{m-1}{2}}\sum_{x_{1:m} \in \mathcal{X}_{1:m}}|x_{1:m}\rangle,
\end{equation*}
where $\mathcal{X}_{1:m}=\{x_{1:m}\in \mathbb {F}^{m}_2|\sum^m_{i = 1}x_i = 0\} $. This allows $m-1$ communications devices to securely transmit $m-1$ bits $Y_1, \cdots, Y_{m-1}$ and compute a parity bit $Y_{m} = Y_1\oplus\cdots\oplus Y_{m-1}$. Such a computation has practical uses in error correction \cite{hamming1950error}.

To measure the amount of secrecy in these joint probabilities, a secrecy monotone obeys (1) semi-positivity, (2) vanishing on product probability distributions, (3) monotonicity under local operations, (4) monotonicity under public communication, (5) additivity, and (6) continuity; see the formal definition in \cite{cerf2002multipartite}. 
As one of the secrecy monotones, the total correlation of a given random variable $X_{1:m}$ is written as 
\begin{align}
C(X_{1:m}) 
= \sum_{i=1}^mH(X_i)-H(X_{1:m})\label{eq:tc},
\end{align}
where $X_{i}$ is the observed variables for subsystem $i \in \mathcal{M}: = \{1,2,\cdots,m\}, m \in \mathbb{N}_+$.  When $m = 1$, we have $C = 0$.  When $m = 2$, total correlation recovers mutual information $I(X_1;X_2)$. Cerf et al. also provide a quantum generalization of secrecy monotone for a quantum state $\rho$, written as 
\begin{equation}\label{eq:tcq}
C_{\text{er}}(\rho) := \sum_{i = 1}^mS(A_i) - S(A_1\cdots A_m),
\end{equation}
where $S(A_i)$ is the von Neumann entropy of subsystem $A_i, i \in \mathcal{M}$, and the whole quantum system is written as $A_1\cdots A_m$.  Groisman et al. claim in \cite{groisman2005quantum} that  $C_{\text{er}}$ measures the total amount of correlations, which includes the amount of quantum and classical correlations and equals the amount of randomness to decorrelate the state. 

By considering the classical and quantum secrecy monotones of the above classical and quantum correlations, we observe that there are resources useful for an information-theoretic task that can be captured by $C_{\text{er}}$, but not by $C$. For $P^{(1)}_{X_{1:m}}$ and $|\psi_1\rangle$, $C = m-1$ and $C_{\text{er}} = m$. The total correlation corresponds to the $m-1$ received bits. The extra bit in $C_{\text{er}}$ is considered as the amount of quantumness in \cite{groisman2005quantum}, which is used to realize quantum phenomena such as non-locality and is therefore not considered as an information-theoretic resource.  Let $P^{(3)}_{X_{1:m}}$ be the measurement outcome of $|\psi_3\rangle$ in the computational basis. We have $C = 1$ and $C_{\text{er}} = m$ for $P^{(3)}_{X_{1:m}}$ and $|\psi_3\rangle$. The total correlation still corresponds to the received bit for the distributed computing problem in Fig.~\ref{fig:LatentStr3}.
 To give an interpretation of the extra $m-1$ bits in $C_{\text{er}}$, we perform the decorrelation process in \cite{groisman2005quantum} to $|\psi_2\rangle$, which can be easily extended to $|\psi_3\rangle$. We apply one of two unitary transformations
$\mathbbm{1}$ or $\sigma_z$ with equal probability on the first qubit and obtain
\begin{align}
    \rho =& \frac{1}{4}|0\rangle \langle 0|\otimes (|00\rangle+|11\rangle) (\langle 00|+\langle 11|)\nonumber\\ &+ \frac{1}{4}|1\rangle \langle 1 |\otimes(|01\rangle+|10\rangle) (\langle 01|+\langle 10|)\nonumber,
\end{align}
We further apply $\mathbbm{1}$ or $\sigma_x$ with equal probability on the first qubit and obtain
\begin{equation*}
    \frac{1}{2}\mathbbm{1} \otimes \frac{1}{4}[(|00\rangle+|11\rangle) (\langle 00|+\langle 11|)+(|01\rangle+|10\rangle) (\langle 01|+\langle 10|)].
\end{equation*}
Now the first qubit is disentangled with the last two qubits. We observe that the state of the last two qubits is the reduced density operator of $A_2A_3$ and a mixture of two maximally entangled states, i.e.,
\begin{equation}
    \frac{1}{4}[(|00\rangle+|11\rangle) (\langle 00|+\langle 11|)+(|01\rangle+|10\rangle) (\langle 01|+\langle 10|)].\label{eq:ap2}
\end{equation}
By applying $\mathbbm{1}$ or $\sigma_z$ with equal probability on the second qubit, we get the disentangled three qubits, i.e., $\frac{1}{2}\mathbbm{1} \otimes \frac{1}{2}\mathbbm{1} \otimes \frac{1}{2}\mathbbm{1}$. In this process, we used three bits of randomness to disentangle $|\psi_2\rangle$. The amount of randomness is equal to $C_{\text{er}}(|\psi_2\rangle\langle\psi_2|)$. One bit of randomness is used in \eqref{eq:ap2} to pair quantum states $|00\rangle$ with $|11\rangle$, and $|01\rangle$ with $|10\rangle$ in two subsystems. 
Consequently, $A_1$ can further employ one bit of randomness to establish the deterministic relationship $X_3 = X_1 \oplus X_2$ with the other two subsystems. However, the measurement outcomes on $A_2A_3$ are independent and $C(P^{(2)}_{X_{2:3}}) = 0$. 
We consider that most of the resources in $|\psi_3\rangle$ are used to coordinate the measurement outcomes on different marginals, in a form similar to \eqref{eq:ap2}, which cannot be measured by total correlation. This leads to the application in the distributed computing problem, where the goal is to coordinate between $m-1$ devices, and all the transmitted $m-1$ bits are necessary, even though the recovered information is only one bit.  To quantify the additional resources that a multipartite entangled state can potentially provide, we introduce a causal model in the next section. 

\section{A causal model for quantifying resources}\label{sec:cas}
In this section, we present our causal model as the source coding side of a  multipartite correlation.  
\begin{definition}[Causal model of multipartite systems]\label{def:cas1}
A causal model of a multipartite system is defined as a directed bipartite graph $G^{(n)} =(\Gamma,V, E)$, where $\Gamma$ and $V$ are two sets of vertices, $E$ is the set of edges and $n\in \mathbb{N}_+$. $\Gamma := \{X_1^n,X^n_2,\cdots, X^n_m\}$ is the set of the observed variables, where each $X_i^n \in \mathcal{X}^n_i$ for $i\in \mathcal{M}$. 
$V := \{\omega_{i,j}\}_{\forall i,j\in \mathcal{M}, i\neq j}$  is the set of the latent variables, where each $\omega_{i,j}$ is a random variable supported on $\Omega_{i,j}$. Note that $\omega_{i,j}$ and $\omega_{j,i}$ refer to the same vertex. $E$ is the set of the edges joining only from $\omega_{i,j}$ to the corresponding $X_i^n$ and $X_j^n$. We denote $\omega_i := \{\omega_{i,j}\}_{j\in \mathcal{M}, j\neq i}$ as the set of the neighbors of $X_i^n$, which is supported on $\Omega_{i} \subset \Omega_{1,i}\times\cdots\times\Omega_{i-1,i}\times\Omega_{i,i+1}\times\cdots\times\Omega_{i,m}$. 

For every node $X^n_i$, a structural equation $X^n_i = f_i : \Omega_i \times \mathcal{U}_i \rightarrow \mathcal{X}^n_i$ holds, 
where $u_i \in \mathcal{U}_i$. 
$\Theta^{(n)} := \{f_{i}(\omega_{i}, u_i)\}_{i\in \mathcal{M}}$ is the set of the structural equations. We define $R_{i,j}(G^{(n)},\Theta^{(n)}) := \log |\Omega_{i,j}|/n$ and $R(G^{(n)},\Theta^{(n)}) := \sum_{i,j}\log |\Omega_{i,j}|/n$ as the rate of the model.
\end{definition}

 \begin{figure}
         \centering
    \includegraphics[width=0.26\textwidth]{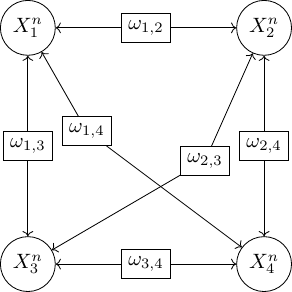}
\caption{A causal model with four observed variables and six latent variables.}
\label{fig:LatentStr}
\end{figure}

The causal model and structural equation in Definition~\ref{def:cas1} were introduced in the general topic of causality (see, e.g., \cite{pearl2009causality}). $\omega_{i,j}$ is the common cause of the two subsystems $i,j$, and $u_i$ is a local term at subsystem $i$. To discuss its interpretation, we provide an example that has been realized in \cite{joshi2020trusted, proietti2021experimental}. 
\begin{example}\label{ex:ca1}
    Consider four entangled photons that are sent to different locations and subsequently measured. The entangled photons display correlations between different locations, and this phenomenon can be described by the causal model in Fig.~\ref{fig:LatentStr}, where $\Gamma := \{X^n_1,X^n_2,X^n_3,X^n_4\}$ and $V:=\{\omega_{1,2}, \omega_{1,3},\omega_{2,3}, \omega_{1,4},\omega_{2,4},\omega_{3,4}\}$. Assume that the measurements are performed on the marginals in their natural order. Following the measurement outcomes $X_1^n$ and $X_2^n$, quantum non-locality is represented as the latent variable $\omega_{1,2}$ shared between marginals 1 and 2. Following $X_3^n$, we obtain $\omega_{1,3}$ and $\omega_{2,3}$. Following $X_4^n$, we obtain $\omega_{1,4}, \omega_{2,4}$ and $\omega_{3,4}$. 
\end{example}

RCCP says that two probabilistic correlated events have a common cause if they are not the cause of each other. In the above example, for each step $i+1$, $\omega_{1:i,i+1}$ is determined once the quantum measurements $\{E_{x_{1:i}}\}$ and $\{E_{x_{i+1}}\}$ are given \cite{van1982charybdis}. Because $\omega_{i+1,i+2:m}$ is determined by both $E_{x_{1:i+1}}$ and the future measurements, from the no-signaling principle, we know that the existing correlation between $X^n_{1:i}$ and $X^n_{i+1}$ is independent with the further measurements given $\{E_{x_{1:i}}\}$ and $\{E_{x_{i+1}}\}$. Therefore, we introduce a condition to characterize the no-signaling condition,  i.e.,
\begin{equation}\label{eq:sw3}
    I(\omega_{1:i};\omega_{i+1,i+2:m}|\omega_{1:i,i+1})=0, \forall i \in \{1,\cdots, m-1\}.
\end{equation} 
We further allow the relationship \eqref{eq:sw3} to tolerate a permutation, written as $\iota:\mathcal{M}\rightarrow \mathcal{M}$, which represents the possible order that the photons are measured. To define the amount of randomness with the above-mentioned properties, we give the next definition. 
\begin{definition}[Compatibility of a causal model]\label{def:cas2}
We say a sequence of models $\{(G^{(n)},\Theta^{(n)})\}_n$ is compatible with $X_{1:m}$, if it satisfies the following conditions:
    \\\textbf{(Condition~1)} \; $\lim_{n\rightarrow \infty}\mathbb{E}_{\Theta^{(n)}}[\tv(P_{X^{n}_{1:m}}, P_{X_{1:m}}^{\otimes n})] = 0$.
   \\\textbf{(Condition~2)} \; There exists a permutation $\iota:\mathcal{M}\rightarrow \mathcal{M}$ such that  $I(\omega_{\iota(1):\iota(i)};\omega_{\iota(i+1),\iota(i+2):\iota(m)}|\omega_{\iota(1):\iota(i),\iota(i+1)})=0, \forall i\in\{1,\cdots,m-1\}$. $I(\{\omega_{i,j}\}_{i,j}; \{u_i\}_i) = 0, I(u_{i};u_{j}) = 0, \forall i,j$.
\end{definition}

Note that there always exists a $\{(G^{(n)},\Theta^{(n)})\}$ that is compatible with an arbitrary $P_{X_{1:m}}$, as we can simply let $\omega_{i,j} = X^{\otimes n}_{1:m}$, $\forall i,j$ and the structural equation as the mapping from $X^{\otimes n}_{1:m}$ to $X^{\otimes n}_{i}$. Condition~1 characterizes the output statistics associated with a given correlation and is defined in the same way as in previous existing literature \cite{han1993approximation,cuff2010coordination,cuff2013distributed,yassaee2014achievability,kurri2021coordination}. 
Conditions~2  can be considered as an information-theoretic extension of the causal order given in \cite{oreshkov2016causal, abbott2016multipartite}, where the amount of randomness between different marginals is further taken into account, and $u_i$ represents the local randomness. We later use the term ``step'' as a synonym for the classical causal order. Based on this definition, we further define resource rate.
\begin{definition}[Resource rate for a classical correlation]\label{def:rr}
Given a sequence $\{(G^{(n)},\Theta^{(n)})\}$, we define $R(\{(G^{(n)},\Theta^{(n)})\}):=\limsup_{n\rightarrow\infty} R(G^{(n)},\Theta^{(n)})$ as its rate. We define $R^*(X_{1:m}):=\inf R(\{(G^{(n)},\Theta^{(n)})\})$ as the resource rate for $P_{X_{1:m}}$, where the infimum is among the sequences compatible with $P_{X_{1:m}}$. A sequence of models $\{(G^{(n)},\Theta^{(n)})\}$ compatible with $P_{X_{1:m}}$ is said to be optimal if $\lim_{n\rightarrow \infty}R(G^{(n)},\Theta^{(n)}) = R^*(X_{1:m})$. 
\end{definition}

\section{Overview of main results}
In Section~\ref{sec:model}, we show the existence of the additional resources that cannot be measured by the classical measure through specific examples. In Section~\ref{sec:cas}, we motivate the causal model from a quantum perspective. In this section, we show how the causal model quantifies the resource in a general correlation. Now, let us provide a semantic definition for the information-theoretic resource under consideration in this paper. Shannon defined in \cite{shannon_mathematical_1948} that the fundamental problem of communication is reproducing a message selected at one point at another point. In adapting this concept from communication to cryptography, the focus shifts from ``reproducing message'' to ``sharing randomness''. Consequently, we define the information-theoretic resource in a given correlation as the amount of randomness at one location that has an effect at another location. Formally, the amount of resources is defined as the resource rate in Definition~\ref{def:rr}. The ``one to another'' condition in the semantic definition corresponds to Condition~2 in Definition~\ref{def:cas2}. To make sure that the randomness has a real effect, we use infimum to make sure it is necessary. To further look into this definition, we show that the proposed causal model characterizes the following three phenomena: 
 \begin{figure}[h]
     \subfloat[\label{fig:g1}]{                
    \includegraphics[width=0.1\textwidth]{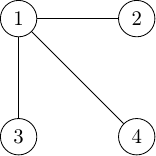}
}
\hfil
\subfloat[\label{fig:g2}]{
    \includegraphics[width=0.1\textwidth]{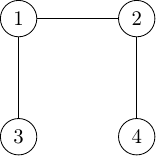}
}
\hfil
\subfloat[\label{fig:g3}]{
    \includegraphics[width=0.1\textwidth]{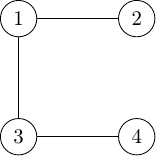}
}
\hfil
\subfloat[\label{fig:g4}]{
    \includegraphics[width=0.1\textwidth]{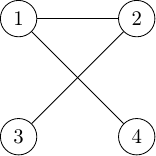}
}
\hfil
\subfloat[\label{fig:g5}]{
    \includegraphics[width=0.1\textwidth]{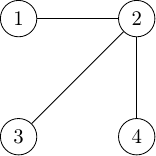}
}
\hfil
\subfloat[\label{fig:g6}]{
    \includegraphics[width=0.1\textwidth]{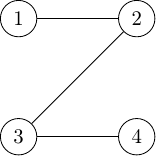}
}
\hfil
\subfloat[\label{fig:g7}]{
    \includegraphics[width=0.1\textwidth]{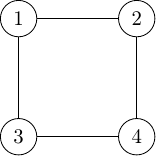}
}

    \caption{Consider the optimal compatible models of a 4-qubit GHZ state with a natural order. Replace the causal forks $X^n_{i} \leftarrow \omega_{i,j}\rightarrow X^n_{j}$ with an undirected edge if there is identical randomness in the latent variables and remove other forks. The resulting graph forms a spanning tree as shown in \protect\subref{fig:g1} to \protect\subref{fig:g6}. For a sub-optimal model, there exist redundant latent variables, forming a circle when added to the spanning tree, as illustrated in \protect\subref{fig:g7}.}
    \label{fig:cas4}
\end{figure}

1. \textit{Identical randomness in different steps}: Consider the example of a 4-qubit GHZ state in \eqref{eq:Q1} where there is identical randomness shared by multiple marginals 
 for broadcast traffic. In a group of users expected to share the same broadcast message, identical randomness of secrecy should cover them with a connected graph, and each pair of users within the group can be connected with a path. As illustrated in Fig.~\ref{fig:cas4}, this is represented as the latent variables with identical randomness. Only one marginal shares the randomness with another at each step in the optimal models. Each spanning tree corresponds to an optimal model.

2. \textit{Identical randomness in one step}: 
Consider the scenario shown in Fig.~\ref{fig:g7}, where marginals 2 and 3 share the same randomness with marginal 4 within one step. One bit of randomness is redundant according to the subadditivity of entropy. Compared to the other cases in Fig.~\ref{fig:cas4}, this case results in an extra edge being added to the spanning tree, thus forming a circle. This redundancy can be eliminated by the Slepian-Wolf coding technique, which forms a tree-like structure for identical randomness.  

3. \textit{Vanishing randomness in one step}: 
Consider the distributed computing problem in Fig.~\ref{fig:LatentStr3}, where two bits of information can be securely transmitted, while only one bit can be recovered from the receiver. 
Similarly, two bits of latent variables need to be non-local, while the structural equation recovers only one bit of randomness at each marginal in the optimal causal model compatible with $P^{(2)}_{X_{1:3}}$. The vanishing randomness is defined as the amount of latent variables that are not recovered by the local structural equations. 

We claim that the vanishing randomness is exactly the reason that a certain amount of resources cannot be measured by \eqref{eq:tc}. This is formally given in the next theorem: 
\begin{theorem}
If there exists a $\{(G^{(n)},\Theta^{(n)})\}$ that is compatible with $P_{X_{1:m}}$ and satisfies
$H(\omega_i|X^n_{i})\leq n\delta, \exists \delta, \forall i$, then $R^*(X_{1:m}) = C(X_{1:m})$ holds necessarily and sufficiently.
Otherwise, we have $R^*(X_{1:m}) > C(X_{1:m})$.
\end{theorem}

In particular, the resource rates for $P^{(1)}_{X_{1:m}}$ and $P^{(3)}_{X_{1:m}}$ are both $m-1$, representing the amount of information that can be securely transmitted. We define the amount of vanishing information as the gap $R^*(X_{1:m}) - C(X_{1:m})$. 

\subsection{Extensions to quantum correlation}
Next, we consider the amount of resources that can be provided by a quantum state. We define quantum total correlation and quantum resource rate in Definitions~\ref{def:qtc} and \ref{def:qms}, respectively. 
\begin{definition}[Quantum total correlation]\label{def:qtc}
Given a density operator $\rho\in \mathcal{S}(\mathcal{H}_1\times\cdots\times\mathcal{H}_m)$, its quantum total correlation is defined as
\begin{equation}\label{eq:qtc}
C_Q(\rho) := \max_{\{E_{x_{j}}\}_{x_j, j\in \mathcal{M}}}C(X_{1:m}),
\end{equation} 
where $\{E_{x_{j}}\}_{x_j, j\in \mathcal{M}}$ is a set of POVMs on the $j$-th marginal and  $P_{X_{1:m}}$ is given by
\begin{equation}\label{eq:measure}
P_{X_{1:m}}(x_{1:m}) = \tr(\rho E_{x_1} \otimes E_{x_2} \otimes\cdots\otimes E_{x_m}).
\end{equation}
\end{definition}

The set of all possible POVMs forms a compact set. The objective function of \eqref{eq:qtc} is continuous from the chain rule and the continuity of entropy. Thus, the maximum exists according to Weierstrass’ extreme value theorem.
\begin{definition}[Quantum resource rate]\label{def:qms}
 Let $P_{X_{1:m}}$ be the measurement outcome  of $\rho$ given by \eqref{eq:measure}. We denote the quantum resource rate as
\begin{equation}
R^*_{Q}(\rho) := \sup_{\{E_{x_{j}}\}_{x_j, j\in \mathcal{M}}}R^*(X_{1:m}).
\end{equation}
\end{definition}

Following the argument in \cite{groisman2005quantum} that $C_{\text{er}}$ is the total amount of the correlations, we obtain the second expression as follows: 
\begin{theorem}\label{th:t2}
For a given pure state $\rho$, and for all $i \in \mathcal{M}$, we have
\begin{equation}\label{eq:qiq}
    C_Q(\rho) \leq R^*_Q(\rho) \leq C_{\text{er}}(\rho) - S(A_i).
\end{equation}
\end{theorem}

Inherited from classical results, the quantum resource rates for $|\psi_1\rangle$ and $|\psi_3\rangle$ are also $m-1$, quantifying the amount of information that can be securely transmitted.

Next, we show the operational meaning of quantum rate from another perspective. When we describe a quantum system, a commonly used model is the quantum circuit. In quantum circuits, entanglement is usually introduced by CNOT gates, which can also be observed from Fig.~\ref{fig:circ1}
. Research has shown that CNOT gate is the only multi-qubit operation required in a universal set of quantum gates to approximate any unitary operation with arbitrary accuracy \cite{nielsen2001quantum}. In the next theorem, we show that the number of CNOT gates is also related to the quantum rate. 

\begin{theorem}\label{th:circ}
    Suppose we have a quantum circuit where a multi-qubit operation is simply a set of CNOT gates. The circuit can be written as a mapping from the input state to the output state, i.e., $|0\rangle^{\otimes m} \rightarrow |\psi\rangle \in \mathbb{C}^{2^m}$. Let $\rho = |\psi\rangle\langle\psi|$. The number of CNOT gates is not smaller than $R^*_Q(\rho)$.
\end{theorem}

\subsection{A hybrid protocol for key distribution}
We next discuss a variant of the omniscient coordination problem with unicast traffic, whereas broadcasting is used in the original problem \cite{kurri2021coordination}. The physical meaning of this variant further provides an information-theoretic analysis for the CKA composition problem via bipartite QKD as mentioned in \cite{murta2020quantum}. 

Now we state our problem settings. Given a joint probability $P_{X_{1:m}}$, our goal is to establish a joint key  $X^n_{1:m}$ with an equivalent key rate to $P_{X_{1:m}}$ (formally proved in the extended version). Let $\alpha_1$ be a QKD server that jointly possesses sufficient bipartite keys with each of the users $j$, where each set of keys is written as $u_j$ and mutually independent. To construct a joint key from the existing bipartite keys, the quantum server and the users follow the following procedure. Let $j$ range from 1 to $m$. At each time step $j$, $\alpha_1$ sequentially sends a classical message $\omega_{1,j} := f_{1,j}(X^n_{1:j-1},u_j)$ to each user $j$, $\omega_{1,j}\in \Omega_{1,j}$. Here, $X^n_{1:j-1}$ refers to the joint key that was established in the previous steps. $X^n_{1:j-1}$ is accessible to $\alpha_1$ because all the previous random variables including $\omega_{1,1:j-1}$ and $u_{1:m}$ are generated by $\alpha_1$. User $j$ then reproduces its part of the joint key $X^n_j := g_{1,j}(\omega_{1,j}, u_j)$ using the received message and its local key.   We define the amount of classical information $\sum_{j = 1}^{m}R_{1,j} := \sum_{j = 1}^{m} \limsup_{n\rightarrow\infty}\frac{1}{n}\log |\Omega_{1,j}|$ as the rate of the above mentioned hybrid systems. We define $R_{\text{hyb}}(X_{1:m}):= \inf \sum_{j = 1}^{m}R_{1,j}$ as the optimal rate of a hybrid system for $P_{X_{1:m}}$, where the infimum is among the sequences of models satisfying the above conditions. We show that:

\begin{theorem}\label{th:hc}
    $R_{\text{hyb}}(X_{1:m})=C(X_{1:m})$.
\end{theorem}

The intuition behind this theorem is to consider the ``source coding side of problem'' introduced in \cite{yassaee2014achievability}. It points out that the sources in network information theory problems can be divided into message and pre-shared randomness. The proposed causal model unfolds multipartite correlations with two conditions, which gives bipartite connections that follow Shannon's semantic definition and include CKA and secret sharing as special cases. In the bipartite connection with common causes, we note that the vanishing randomness corresponds to the pre-shared randomness in the source coding side of problem, which can be eliminated using a random codebook, i.e., the bipartite QKD in our problem. 
The achievability of Theorem~\ref{th:hc} can be written as the pseudocode in Algorithm~\ref{al:sc}. For binary CKA, Line~\ref{line:1} becomes $\|X^n_{1} - X^n_{j}[k]\|\leq n\epsilon_j$ with sufficient small $\epsilon_j$. The provided coding scheme has time and space complexity $\mathcal{O}(n\sum^m_{j=2}2^{nR_{1,j}})$, and uses only $X^n_j[\omega_{1,j}]$ among all the buffered key $u_j:= X^n_{j}[0:2^{nR_{1,j}}]$ when coding the classical information $\omega_{1,j}$. Therefore, the unused key remains private and can be used later, which improves the resource efficiency of the coding scheme. The security against eavesdropping is from the fact that the bipartite keys are quantum secure. 

\begin{algorithm}[H]
  \caption{Key distribution with a centralized server}\label{al:sc}
    \textbf{Input:} $X^n_{1}, X^n_{2}[0:2^{nR_{1,2}}],
    \cdots, X^n_{m}[0:2^{nR_{1,m}}]$\\
  \textbf{Output:} $\omega_{1,2:m}$
  \begin{algorithmic}[1]
  \FOR{$j = 2;\ j \leq m;\ j = j + 1$}
    \FOR{$k = 0;\ k < 2^{nR_{1,j}};\ k = k + 1$}
        \IF{$(X^n_{1:j-1},X^n_{j}[k])$ meets certain criteria}\label{line:1}
        \STATE $\omega_{1,j} = k$
        \STATE $X^n_{j} = X^n_{j}[\omega_{1,j}]$
        \STATE \textbf{break}
        \ELSIF{$k = 2^{nR_{1,j}}-1$} \RETURN error
        \ENDIF
    \ENDFOR
  \ENDFOR
  \RETURN $\omega_{1,2:m}$

  \end{algorithmic}
\end{algorithm}

For proofs and further details in this section, we refer the reader to the extended version of this manuscript \cite{wang2024causal}. 

\section*{Acknowledgment}
The authors were supported by the German Science Foundation (DFG) as part of SPP 2378 (Resilient Worlds: project number 503691052, ResCTC) and the Federal Ministry of Education and Research of Germany (BMBF) in the programme of ``Souverän. Digital. Vernetzt.'' Joint project 6G- RIC, project identification number 16KISK020K.

\bibliographystyle{IEEEtran}
\bibliography{apssamp}

\end{document}